\documentclass[conference]{IEEEtran}
\usepackage{arydshln}
\IEEEoverridecommandlockouts
\usepackage[font=small,labelfont=bf]{caption}

\usepackage{mwe}
\usepackage{graphicx} % For including images
\usepackage{adjustbox} % To adjust the size of images
\usepackage{booktabs} % To adjust the size of images
\usepackage{lipsum}  
\usepackage{tikz}
\usepackage{multirow}
\usepackage{acro}
\usepackage{cite}
\usepackage{amsmath,amssymb,amsfonts}
\usepackage{algorithmic} 
\usepackage{textcomp}
\usepackage{xcolor}
\usepackage{hyperref}
\usepackage{cleveref}
\usepackage{siunitx}

\hypersetup{
    colorlinks=true,
    linkcolor=blue,
    filecolor=magenta,      
    urlcolor=magenta,
    citecolor=blue
}
\DeclareAcronym{ai}{short=AI, long= Artificial Intelligence}
\DeclareAcronym{ann}{short=ANN, long= Artificial Neuron Network}
\DeclareAcronym{cnn}{short=CNN, long=Convolutional Neuron Network} 
\DeclareAcronym{dvs}{short=DVS, long=Dynamic Vision Sensor}  
\DeclareAcronym{if}{short=IAF, long= Integrate And Fire}
\DeclareAcronym{lstm}{short=LSTM, long= Long-Short-Term-Memory} 
\DeclareAcronym{mac}{short=MAC, long= Multiply and Accumulate} 
\DeclareAcronym{nir}{short=NIR, long=Near-Infra Red (Illumination)} 
\DeclareAcronym{nas}{short=NAS, long=Neural Architecture Search} 
\DeclareAcronym{snn}{short=SNN, long= Spiking Neuron Network} 
\DeclareAcronym{soc}{short=SoC, long=System-on-Chip} 
\DeclareAcronym{speck}{short=Speck, long= Synsense Speck}  
\DeclareAcronym{yolo}{short=YOLO, long=You Only Look Once} 
\DeclareAcronym{iou}{short=IoU, long=Intersection Over Unit} 

\def\BibTeX{{\rm B\kern-.05em{\sc i\kern-.025em b}\kern-.08em
    T\kern-.1667em\lower.7ex\hbox{E}\kern-.125emX}}
\begin{document}
 
\title{Sub-Millisecond Event-Based Eye Tracking on a Resource-Constrained Microcontroller \\
\thanks{\IEEEauthorrefmark{1} These authors contributed equally to this work

This work was funded by the Innosuisse (103.364 IP-ICT) and Swiss National Science Foundation (Grant 219943).\\
Corresponding authors: \{marco.giordano, pietro.bonazzi\}@pbl.ee.ethz.ch}
}
\author{
    \IEEEauthorblockN{Marco Giordano\IEEEauthorrefmark{1}, Pietro Bonazzi\IEEEauthorrefmark{1}, Luca Benini, Michele Magno}
    
    \IEEEauthorblockA{ETH Zürich, Zürich, Switzerland} 
}

\maketitle

\begin{abstract}
This paper presents a novel event-based eye-tracking system deployed on a resource-constrained microcontroller, addressing the challenges of real-time, low-latency, and low-power performance in embedded systems. The system leverages a \ac{dvs}, specifically the DVXplorer Micro, with an average temporal resolution of \SI{200}{\micro\second}, to capture rapid eye movements with extremely low latency. The system is implemented on a novel low-power and high-performance microcontroller from STMicroelectronics, the STM32N6. The microcontroller features an \SI{800}{\mega\hertz} Arm Cortex-M55 core and AI hardware accelerator, the Neural-ART Accelerator, enabling real-time inference with milliwatt power consumption. The paper propose a hardware-aware and sensor-aware compact \ac{cnn} optimized for event-based data, deployed at the edge, achieving a mean pupil prediction error of 5.99 pixels and a median error of 5.73 pixels on the Ini-30 dataset. The system achieves an end-to-end inference latency of just \SI{385}{\micro\second} and a neural network throughput of 52 \ac{mac} operations per cycle while consuming just \SI{155}{\micro\joule} of energy. This approach allows for the development of a fully embedded, energy-efficient eye-tracking solution suitable for applications such as smart glasses and wearable devices.
\end{abstract}


\section{Introduction}

With the rapid advancement of computer vision, machine learning, and consumer electronics, eye‑tracking has emerged as a modality across numerous domains, including high bandwidth human-computer interfacing, virtual and augmented reality (VR/AR)\cite{adhanom2023eye}, cognitive neuroscience\cite{wolf2023eye}, and assistive technologies\cite{vessoyan2023scoping}. By monitoring gaze patterns and eye movements, eye tracking enables intuitive user interfaces, enhances immersive experiences, and provides insights into cognitive processes \cite{yang2023wearable}. Data must also be processed on the device, since continuous data transmission to external servers poses challenges for integration into compact, battery-powered devices, where energy efficiency and low-latency processing are paramount\cite{giordano2022survey}. 
 As the demand for real-time, portable, and energy-efficient eye-tracking solutions grows\cite{chen20233et, Bonazzi_2024_CVPR}, especially in wearable devices like smart glasses and VR headsets, the limitations of traditional eye-tracking systems become increasingly apparent\cite{angelopoulos2020event}.

\begin{figure}[t]
    \centering
    \includegraphics[width=\linewidth]{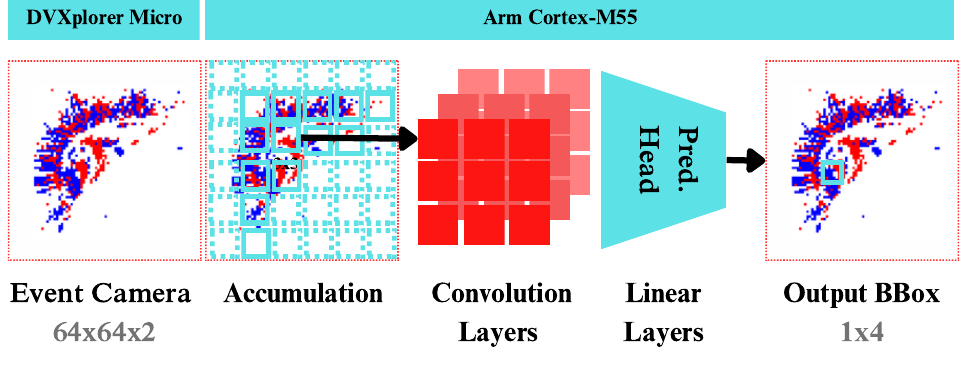}
    \caption{Overview of the proposed event-based eye tracking system.}
    \label{fig:architecture}
\end{figure}

Conventional eye-tracking technologies predominantly rely on frame-based cameras\cite{ryan2021real}, often coupled with infrared illumination\cite{ryan2021real}, to capture images of the eye at fixed intervals. While effective in controlled environments, these systems often suffer from high latency\cite{ryan2021real}, substantial power consumption\cite{Bonazzi_2024_CVPR}, and sensitivity to varying lighting conditions\cite{ryan2021real}.

Event-based cameras are an emerging and promising technology improving both latency and energy efficiency \cite{zhao2023ev}. Unlike traditional cameras that capture entire frames at regular intervals, event-based sensors asynchronously detect changes in luminance at each pixel, producing a stream of events with microsecond temporal resolution. This approach results in reduced data redundancy\cite{gallego2020event}, lower latency\cite{gallego2020event}, and enhanced dynamic range\cite{gallego2020event}, making event-based cameras particularly suited for capturing rapid eye movements and operating under diverse lighting conditions \cite{zhao2023ev}.

Despite these advantages, integrating event-based eye tracking systems into wearable resource-constrained environments, such as smart glasses, remains an open challenge \cite{Bonazzi_2024_CVPR}. In fact, due to the stringent power and memory budgets of wearable processors, the microsecond data throughput generated by \ac{dvs}, the need for algorithms that can process sparse asynchronous events in real time\cite{covi2021adaptive}, and, crucially, the current lack of mature end-to-end hardware software pipelines that run entirely on embedded or neuromorphic SoCs, existing solutions often depend on high-performance processors or external computing resources, which are unsuitable for wearable applications \cite{chen20233et}. Thus, the need for a self-contained, efficient, and accurate eye-tracking system that leverages the benefits of low-power event-based vision while operating within the constraints of embedded hardware is evident. To meet this challenge, tightly integrated algorithm hardware co-design will require combining low footprint models with ultralow power accelerators, supported by open embedded benchmarks and system‑level evaluations to guarantee real‑time performance, energy efficiency, and reproducibility \cite{Bonazzi_2024_CVPR, iddrisu2024survey}.

This paper presents an end-to-end event-based eye-tracking system designed for deployment on a resource-constrained microcontroller, \Cref{fig:architecture}, that can be fully embedded in a wearable device supplied by a few hundred mAh battery. Our approach utilizes \ac{dvs} to capture high temporal resolution eye movement data, which is processed by the proposed compact \ac{cnn} designed and optimized for event-based inputs. The CNN is deployed on a novel AI processor, the STM32N6 microcontroller, featuring an Arm Cortex-M55 core and a Neural-ART Accelerator. We demonstrated that our model running on the STM32n6 enables real-time inference with sub \SI{1}{\milli\second} latency using only 155 uJ per inference. By performing all processing on-device, our system eliminates the need for external computation or wireless data transmission, thereby reducing latency and energy usage.

Our contributions are as follows:
\begin{itemize}
    \item Design and Implementation: We develop an end-to-end event-based eye-tracking system that operates entirely on a microcontroller, demonstrating the feasibility of deploying neural-network based eye tracking in resource-limited environments.
    \item Model Optimization: We introduce a lightweight CNN tailored for event-based data, achieving high accuracy in pupil localization at low computational complexity.
    Moreover, we quantize the network to 8-bit integers with minimal loss, in ensuring accuracy and efficient execution on the STM32N6
    \item Comprehensive Evaluation: We assess our system's performance using the Ini-30 dataset, achieving an inference latency of 204 us, while maintaining near-SOTA performance (+2px centroid error) highlighting its suitability for real-time applications.
\end{itemize}

The remainder of this paper is organized as follows. Section II reviews existing approaches to eye tracking, emphasizing the limitations of traditional frame-based systems and highlighting recent advancements in event-based vision and embedded processing. Section III details our approach to event-based eye tracking, including input representation, neural network architecture, quantization techniques, and training procedures. Section IV describes the hardware components of our system, focusing on the STM32N6 microcontroller and the DVXplorer Micro event-based camera, and explains how they are integrated for efficient on-device processing. Section V presents the performance metrics of our system, including accuracy, latency, and energy consumption, and compare our results with existing methods to demonstrate the efficacy of our approach. Section VI concludes with a summary of our contributions and discusses potential directions for future research in event-based eye tracking on resource-constrained devices.

By addressing the challenges of integrating event-based eye tracking into embedded systems, our work paves the way for the development of energy-efficient, low-latency eye-tracking solutions suitable for next-generation wearable devices.

\begin{table}[h]
\centering
\caption{A breakdown of the network complexity evaluated without temporal bins.}
\label{table:mac-params}
\small
\begin{tabular}{c|c|c}
\toprule
Method &  MAC $\downarrow$ & Parameters $\downarrow$  \\
\midrule 
3ET \cite{chen20233et}  & 54.95M & 425.36k \\
Retina \cite{Bonazzi_2024_CVPR} &  3.53M & 52.78k \\  
Ours &  10.2M & 59.57k \\  
\bottomrule
\end{tabular}
\end{table}

\section{Related Work}

Event-based vision has emerged as a promising alternative to traditional frame-based imaging for various computer vision tasks, including eye tracking \cite{iddrisu2024survey, chen2025event}. Event cameras offer advantages such as high temporal resolution, low latency, and high dynamic range, making them well-suited for capturing rapid eye movements.

Early work in event-based eye tracking explored hybrid systems combining frame-based and event data. Angelopoulos et al. \cite{angelopoulos2020event} introduced a hybrid near-eye gaze tracking system achieving high update rates and computational efficiency. Stoffregen et al. \cite{stoffregen2022event} presented the first fully event-based and model-based glint tracker, demonstrating robust tracking at 1 kHz with low power consumption. Zhao et al. \cite{zhao2023ev} further advanced hybrid approaches with a matching-based pupil tracking method.

More recent research has focused on leveraging the unique characteristics of event data with advanced neural network architectures. Chen et al. \cite{chen20233et} proposed a sparse change-based convolutional LSTM model to reduce computational load while preserving accuracy in event-based eye tracking. Wang et al. \cite{wang2024mambapupil} introduced a bidirectional selective recurrent model to handle the diversity and abruptness of eye movements, enhancing tracking robustness.

To address the need for efficient real-time eye tracking, particularly in resource-constrained environments, researchers have explored specialized hardware and processing techniques. Tan et al. \cite{tan2025arvr} developed a DVS-based processor for low-latency and high-accuracy eye tracking in AR/VR devices. Schärer et al. \cite{scharer2024electrasight} proposed smart glasses with fully onboard non-invasive eye tracking, highlighting the potential for integrated solutions.

\ac{snn} have gained attention for their energy efficiency and event-driven nature, making them suitable for processing event camera data for near-eye pupil-tracking \cite{Bonazzi_2024_CVPR} and gaze-tracking \cite{groenen2025gazescrnn}. 

Applications leveraging event-based eye tracking are also being explored. Zhao et al. \cite{zhao2025evgaze} proposed an eye authentication system based on micro eye movements captured by event cameras, emphasizing the potential for secure user verification in low-power devices. Ren et al. \cite{ren2025temporal} investigated the temporal dynamics of eye movements to improve the accuracy and speed of event-based eye trackers. Ding et al. \cite{ding2024facet} introduced FACET, a fast and accurate event-based eye-tracking model for extended reality applications and Sen et al. \cite{sen2024eye} developed a low-latency adaptive event slicing method for fine-grained eye tracking, suitable for edge-based solutions.

\textbf{Gaps in the Literature and Motivation:} While significant progress has been made in event-based eye tracking, a notable gap exists in the literature concerning the deployment of these advanced algorithms on resource-constrained microcontrollers. Existing research primarily focuses on algorithmic advancements and performance evaluation on desktop-grade hardware or specialized processors. The unique challenges \cite{ bonazzi2023tinytracker} associated with deploying event-based eye-tracking models on microcontrollers with limited memory, computational power, and energy budgets, remain largely unexplored.
\section{Methodology} 

\subsection{Event-Based Eye Tracking}

\subsubsection{Input Representation}

\begin{figure}[t]
    \centering
    \includegraphics[width=\linewidth]{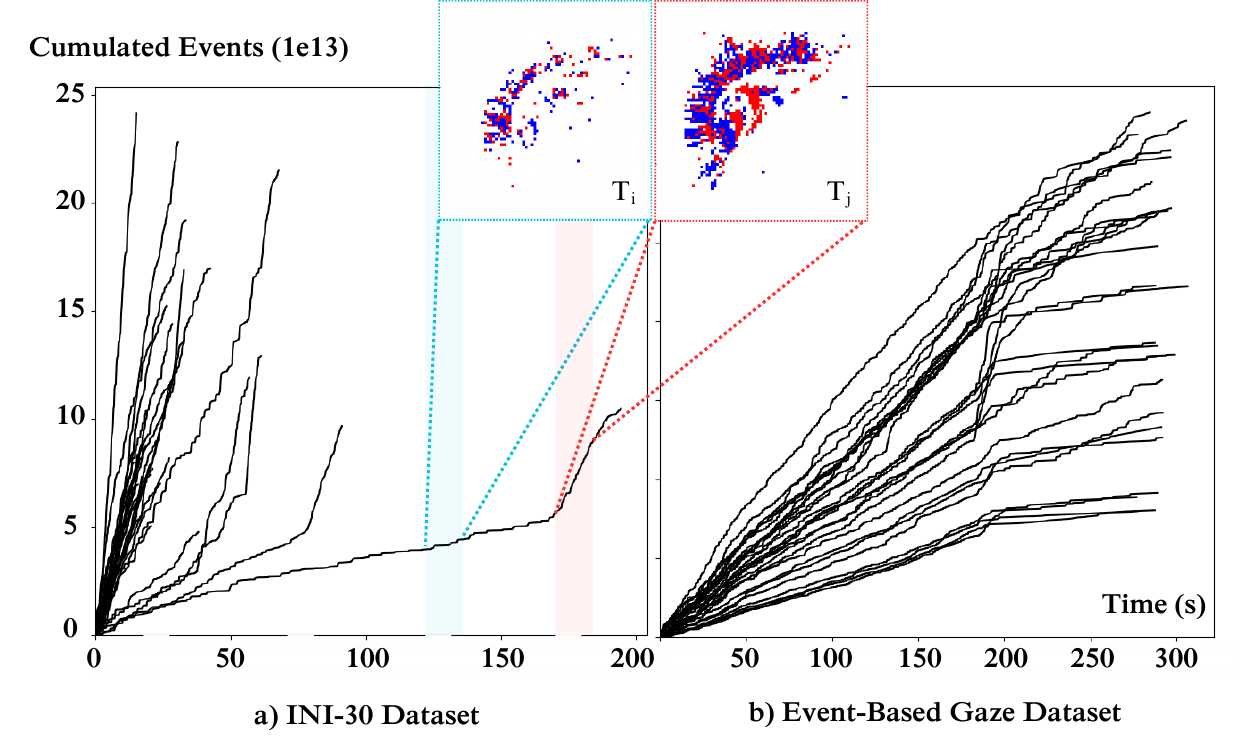}
    \caption{Temporal evolution of the number of events over time in a sample video sequence in Ini-30 \cite{Bonazzi_2024_CVPR} and in the Event-Based Gaze Dataset \cite{angelopoulos2020event}. The event frames highlight the variable density of events across equal time windows ($T_i = T_j$).}
    \label{fig:temporal_stats}
\end{figure}

The input to the model consists of event frames generated from a \ac{dvs}. The non-uniform temporal distribution of events produced by the \ac{dvs} used in the Ini-30 Dataset \cite{Bonazzi_2024_CVPR} is illustrated in \Cref{fig:temporal_stats}. As seen in Panel A, the number of events fluctuates significantly throughout the sequence due to changes in motion dynamics and scene activity. This variability motivates the choice of slicing the event stream based on a fixed number of events rather than fixed time intervals. By maintaining a constant number of events per frame (e.g., 1000), the resulting input representation ensures consistent information density across samples, which facilitates stable model training and improves generalization. The duration of the accumulation is highly depended on the quantity of events generated by the users. We refer to \cite{Bonazzi_2024_CVPR} for statistics related to Ini-30. The events are discretized into a single time bin and encoded as a 2-channel image of size $64 \times 64$ (corresponding to positive and negative event polarities). Various data augmentations such as spatial scaling (factor 0.25) and center cropping are applied to improve generalization.

\subsubsection{Network Architecture}

The proposed eye tracking model is a compact convolutional neural network designed for low-resolution event-based input, with an input resolution of $64 \times 64$ pixels and a single channel. The architecture is inspired by Retina \cite{Bonazzi_2024_CVPR} and optimized for low-latency, low-power inference. It comprises six convolutional blocks, each consisting of convolution, normalization, and activation layers, followed by a fully connected prediction head that outputs normalized bounding box coordinates through a sigmoid activation. 

The input to the network is a tensor of shape $64 \times 64 \times 1$. The first convolutional layer applies $16$ filters of size $5 \times 5$ with stride $2$ and padding $1$, followed by batch normalization and a ReLU activation function. The second convolutional layer increases the number of output channels to $64$ using $3 \times 3$ filters with stride $1$ and padding $1$, followed by batch normalization and ReLU. The third convolutional layer reduces the number of channels to $16$, using the same kernel size, stride, and padding, and is again followed by batch normalization and ReLU. The fourth convolutional block repeats the configuration of the third layer (i.e., $16$ filters of size $3 \times 3$, stride $1$, padding $1$), followed by batch normalization and ReLU. After this block, spatial resolution is reduced using average pooling with a $2 \times 2$ kernel and stride $2$. The fifth block consists of a convolution with $8$ filters of size $3 \times 3$, stride $1$, and padding $1$, followed by batch normalization, ReLU, and another $2 \times 2$ average pooling operation with stride $2$. The sixth convolutional block restores the channel dimension to $16$, again using $3 \times 3$ filters with stride $1$ and padding $1$, followed by batch normalization, ReLU, and a final $2 \times 2$ average pooling layer.

The output feature maps are flattened into a one-dimensional vector and passed through a fully connected layer with $128$ output units and a ReLU activation. The final prediction head is a fully connected linear layer with $4$ output values, corresponding to the normalized bounding box parameters $(x_\text{center}, y_\text{center}, w, h)$. Finally, a sigmoid activation is applied to constrain the outputs to the interval $[0, 1]$.

\subsubsection{Quantization}

To reduce computational cost and enable deployment on resource-constrained edge devices, we apply static post-training quantization to the trained model\cite{jacob2018quantization}. The method relies on quantizing both weights and activations from 32-bit floating-point precision to 8-bit integers using the Quantize-DeQuantize (QDQ) format. A representative dataset, derived from the training distribution, is used to calibrate the model’s activation ranges. This calibration process ensures that the quantized model retains the dynamic characteristics of the original floating-point model. We use the MinMax calibration method\cite{vanhoucke2011improving}, which captures the range of activations by observing the minimum and maximum values over the calibration set. Per-channel quantization is applied to the weights to enhance accuracy, while per-tensor quantization is used for activations. The approach also includes range reduction to ensure compatibility with hardware accelerators. Symmetric quantization is used for weights to simplify implementation on integer arithmetic engines, whereas asymmetric quantization is used for activations to better adapt to data variability. The resulting quantized model retains inference accuracy while significantly reducing model size and computational requirements, making it suitable for real-time, on-device eye tracking.

\subsubsection{Training Details}

The network is trained using the Adam optimizer \cite{kingma2017adammethodstochasticoptimization} with a learning rate of 0.001 and a batch size of 32. Training is conducted for one epoch, and a step learning rate scheduler is employed to adjust the learning rate during training. The loss function comprises two main components: a bounding box \ac{iou} loss weighted by 7.5 and a Euclidean point distance loss weighted by 1.0. The model predicts directly a single bounding box, the center of which is compared to the ground truth to evaluate the model's precision.

\subsection{System Design}

\subsubsection{STM32N6}

The STM32N6 is a high-performance microcontroller from STMicroelectronics, engineered specifically for edge AI and TinyML applications. It features an \SI{800}{\mega\hertz} Arm Cortex-M55 core with Helium vector extensions, enhancing digital signal processing capabilities crucial for real-time inference tasks.

A standout component is the integrated Neural-ART Accelerator, ST's proprietary neural processing unit (NPU), clocked up to 1GHz. This NPU enables efficient on-device execution of complex neural networks, as demonstrated by our eye-tracking usecase.

The microcontroller includes 4.2MB of contiguous embedded RAM, facilitating the handling of data-intensive AI workloads. Its architecture supports high-speed external memory interfaces like OCTOSPI and hexa-SPI, providing flexibility for expanded storage needs.

For prototyping, we employed the STM32N6 Nucleo-144 development board (MB1940), which features the STM32N657X0H3Q microcontroller. In our implementation, we opted to utilize only the microcontroller's embedded SRAM for both code and data storage, foregoing the use of external flash memory.

The STM32N6 is supported by ST's Edge AI Suite, including tools like STM32Cube.AI and the STM32 model zoo, streamlining the deployment of machine learning models on embedded systems.

\subsubsection{DVXplorer Micro}

To capture rapid eye movements with high temporal precision, we employed the DVXplorer Micro, a compact event-based vision sensor developed by iniVation. Unlike traditional frame-based cameras, the DVXplorer Micro operates on an event-driven paradigm, asynchronously detecting changes in luminance at each pixel. This approach significantly reduces data redundancy and latency, making it particularly suitable for real-time eye-tracking applications.

The DVXplorer Micro features a $640 \times 480$ pixel array, which gets accumulated into a $64 \times 64$ pixel grid during preprocessing, with a pixel pitch of \SI{9}{\micro\meter}, providing a temporal resolution of \SI{200}{\micro\second} and typical latency below 1 ms. Its dynamic range reaches up to \SI{110}{\decibel}, enabling operation under varying lighting conditions. The sensor can handle up to 450 million events per second, ensuring responsiveness to rapid eye movements. Additionally, the integrated 6-axis inertial measurement unit (IMU) offers gyroscope and accelerometer data at sampling rates up to \SI{8}{\kilo\hertz}, facilitating sensor fusion and motion compensation.

\begin{figure}[b]
    \centering
    \includegraphics[width=\linewidth]{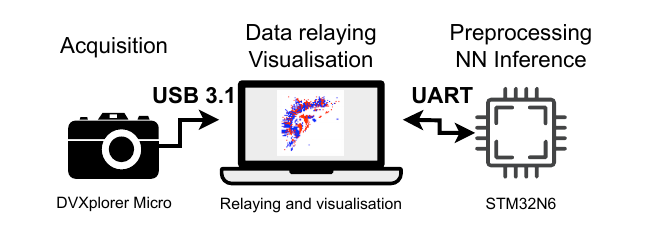}
    \caption{Illustration of the system used to demo the eye-tracking application.}
    \label{fig:system}
\end{figure}

The camera's compact dimensions (24 × 27.5 × 29.7 mm) and lightweight design (\SI{16}{\gram} without lens) make it suitable for embedded applications. It utilizes an S-mount (M12) lens interface and connects via a USB 3.1 Type-C port, which supplies both data and power. In our setup, illustrated in Figure \ref{fig:system}, the DVXplorer Micro camera is connected to a host PC via USB 3.1, where a Docker container running iniVation's DV software accesses the raw event data. The PC acts as a conduit, relaying this data over a 4 Mbps UART interface to the STM32N6 microcontroller. The STM32N6 performs all event accumulation, neural network inference using its Neural-ART Accelerator, and postprocessing. Final pupil coordinates are sent back to the PC for visualization, making the system a fully embedded eye-tracking solution with the PC serving only as a relay and display interface.

% \subsubsection{Application}

% In our experimental setup, the DVXplorer Micro camera was connected to a host PC via USB 3.1. Due to the proprietary nature of the camera's USB protocol, direct interfacing with the STM32N6 microcontroller was not feasible. To address this, we utilized a Docker container running iniVation's DV software suite on the host PC, which facilitated access to the raw event data generated by the camera.

% The host PC acted solely as a conduit, relaying the unprocessed event data to the STM32N6 microcontroller over a UART interface configured at a baud rate of 4 megabits per second. This UART communication was established over a USB connection, ensuring reliable and high-speed data transfer between the host PC and the microcontroller.

% Upon receiving the event data, the STM32N6 microcontroller performed all subsequent processing steps. This included accumulating the events, executing a neural network inference using its integrated Neural-ART Accelerator, and postprocessing to interpret the results. The final eye-tracking data, specifically the x and y coordinates of the pupil, were then sent back to the host PC for visualization and analysis.

% This architecture effectively constitutes a fully embedded, end-to-end eye-tracking solution, with the host PC serving only as a data relay and visualization platform.
\section{Experimental Results} 
\begin{figure}[t]
    \centering
    \vspace{0.55cm}
    \includegraphics[width=0.95\linewidth,trim=0cm 0cm 0cm 0.8cm,clip]{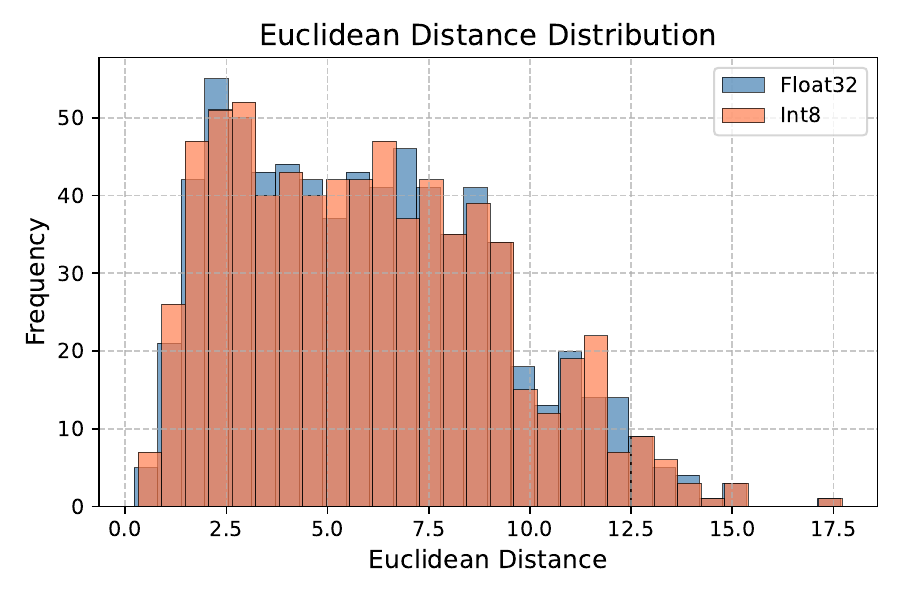}
    \vspace{-0.15cm}
    \caption{Distribution of errors in Ini-30 \cite{Bonazzi_2024_CVPR} validation set for full precision and quantized models.}
    \vspace{0.3cm}
    \label{fig:distribution_errors}
\end{figure}

\subsection{Benchmark Comparison} 

Similar to Retina \cite{Bonazzi_2024_CVPR}, we followed a leave-two-participants-out cross-validation protocol for the Ini-30 dataset. The results from the evaluation are summarized in \Cref{fig:distribution_errors}, showing the performance of both the full precision and quantized models.

In the results, both models exhibit similar performance. The full-precision model shows a mean centroid error of 5.99 pixels, with a median of 5.73 pixels. The quantized model, on the other hand, has a mean centroid error of 5.98 pixels and a median of 5.76 pixels. The differences between these two models are minimal, suggesting that quantizing the model to 8-bit does not significantly impact accuracy, as both models showing comparable performance in terms of error distribution, with only slightly higher interquartile range (5.090 vs 5.123, 0.7\%) for the quantized version.

\begin{table*}[h]
\centering
\caption{Latency, energy, and performance metrics of the preprocessing and INT8 neural network inference on the STM32N6. The preprocessing accounts for 1500 events.}
\label{table:combined-performance}
\small
\begin{tabular}{c|c|c|c|c}
\toprule
Task & Latency $\downarrow$ & Energy $\downarrow$ & Throughput $\uparrow$ & Median Dist. $\downarrow$ \\
\midrule 
Preprocessing  & \SI{119}{\micro\second} & \SI{69}{\micro\joule} & -- & -- \\
Loading        & \SI{62}{\micro\second}  & \SI{35}{\micro\joule} & -- & -- \\
NN (INT8)      & \SI{204}{\micro\second} & \SI{155}{\micro\joule} & 52 MAC/Cycle & 5.76px \\
\bottomrule
\end{tabular}
\end{table*}

In \Cref{table:mac-params}, we present a comprehensive comparison between our proposed model with novel stateful networks such as Retina \cite{Bonazzi_2024_CVPR} and 3ET \cite{chen20233et}, focusing on the number of parameters and the volume of \ac{mac} operations. Despite having slightly more MAC operations than Retina \cite{Bonazzi_2024_CVPR}, our model demonstrates reduced complexity compared to 3ET\cite{chen20233et}, making it more efficient in terms of both MAC operations and parameter count. Our model falls between Retina\cite{Bonazzi_2024_CVPR} and 3ET\cite{chen20233et} in terms of both MAC operations and parameter count, underscoring its efficiency.

\subsection{System Evaluation}

\Cref{table:combined-performance} provides a comprehensive breakdown of the computational pipeline on the STM32N6, detailing latency, energy consumption, and performance metrics for each processing stage. These results demonstrate that our system meets the stringent real-time requirements of embedded eye tracking, with total end-to-end processing latency well below 1 ms.

The preprocessing step, which accumulates and organizes 1500 \ac{dvs} events into a structured input tensor, takes only \SI{119}{\micro\second}, consuming \SI{69}{\micro\joule}. This latency highlights the efficiency of our event-slicing strategy and the lightweight operations required for spatial discretization and polarity channel separation. The low energy consumption also validates our choice of simple but effective preprocessing operations, which are suitable for battery-powered applications.

The data loading phase, responsible for transferring the preprocessed tensor into inference memory, adds an overhead of \SI{62}{\micro\second} and \SI{35}{\micro\joule}. This step benefits from the high-throughput memory interface and the availability of contiguous on-chip SRAM on the STM32N6, which prevents bottlenecks that often occur in resource-constrained systems during data transfer.

\begin{figure}[t]
    \centering
    \includegraphics[width=\linewidth]{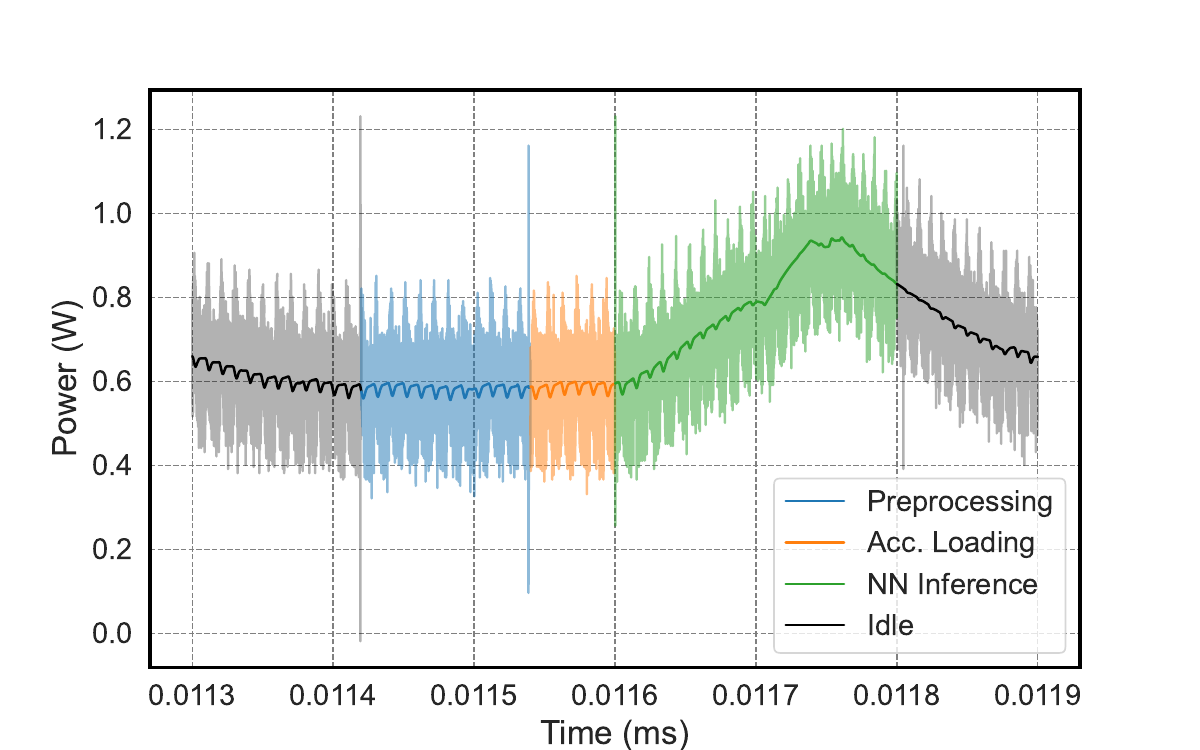}
    \caption{Power profile of preprocessing, neural network loading, and inference. The shaded area represents the raw data, the bolded line is a moving average.}
    \label{fig:power_profile}
\end{figure}

The neural network inference, performed entirely on-device using the Neural-ART Accelerator, constitutes the most computationally intensive stage. Despite this, it completes in just \SI{204}{\micro\second}, while consuming \SI{155}{\micro\joule}. Importantly, this inference is executed on a quantized INT8 model, which achieves a throughput of 52 MACs per cycle, a figure that underscores the hardware acceleration advantages of the STM32N6.

Overall, the total processing pipeline—preprocessing, loading, and inference—requires just \SI{385}{\micro\second}, confirming that our system performs sub-millisecond gaze estimation. This low-latency, low-power profile, combined with competitive inference accuracy, demonstrates that full on-device processing for event-based eye tracking is not only feasible but practical, opening new opportunities for energy-constrained wearable platforms such as smart glasses and AR/VR headsets.

Figure~\ref{fig:power_profile} illustrates the power consumption profile of the system during a single inference cycle, segmented into its three principal stages: preprocessing, accelerator loading, and neural network inference. The shaded regions correspond to raw instantaneous current measurements, while the bold lines represent a moving average for clarity. During the idle phase, the system maintains a baseline current draw of approximately \SI{0.6}{\watt}, primarily due to the microcontroller remaining in an active state to ensure rapid response times. The preprocessing stage (blue) shows a moderate but stable power level, reflecting lightweight CPU-bound operations. In the accelerator loading phase (orange), the current remains relatively constant, indicating minimal computational activity but active memory transfers to prepare the accelerator. Power consumption peaks during the neural network inference phase (green), where the Neural-ART Accelerator is fully engaged. The sharp rise in current in this phase correlates with the increased MAC throughput and high switching activity on the processing elements. Following inference, the system returns to its idle baseline. This profile confirms that the energy cost is heavily concentrated in the inference stage, while the preprocessing and loading stages are significantly less demanding, supporting the system's suitability for energy-efficient intermittent operation.

To estimate the battery lifetime of our system under realistic operating conditions, we consider a \SI{150}{\milli\ampere\hour} lithium-polymer battery, targeting a pair of smart glasses, and a \SI{4000}{\milli\ampere\hour} target a VR headset, both at \SI{3.7}{\volt}, providing approximately \SI{0.56}{\watt\hour} (\SI{2}{\kilo\joule}) and \SI{14.8}{\watt\hour} (\SI{53.3}{\kilo\joule}) of usable energy.

At an inference rate of 120 gaze estimations per second,  to match high-refresh rate displays, and with each full inference pipeline (preprocessing, loading, and INT8 neural inference) consuming \SI{259}{\micro\joule}, the active energy use amounts to \SI{31.1}{\milli\joule} per second.
Additionally, each \SI{8.33}{\milli\second} cycle includes ~\SI{7.95}{\milli\second} of idle time, during which the system is estimated to draw \SI{2}{\milli\watt}, contributing an additional \SI{15.9}{\micro\joule} per cycle, accounting for \SI{1.9}{\milli\joule} per second. This brings the total average energy consumption to ~\SI{33.0}{\milli\joule} per second.

Under these conditions, the system can operate continuously for about a full wake-up day (\SI{16}{h}) in the smart glasses, and over two weeks on the VR headset on a single charge. In comparison, commercial AR/VR headsets typically provide around 2-3 hours of battery life. Matching that runtime would consume only about 0.6\% of our system’s battery capacity to extract eye position from event data. This highlights the energy efficiency and practical viability of our approach for continuous, high-frequency eye tracking in wearable platforms.
\section{Conclusion} 

We presented a fully embedded, event-based eye-tracking system running on a resource-constrained STM32N6 microcontroller, demonstrating that real-time inference is achievable with sub-millisecond latency and minimal energy consumption. By leveraging an optimized quantized CNN and the Neural-ART Accelerator, our solution achieves a median gaze localization error of 5.76 pixels on the Ini-30 dataset with an inference latency of just \SI{204}{\micro\second}. Our design is projected to last more than 16 hours on a fully charged smart-glasses battery and to consume around 1\% of a fully charged VR headset, while running continously at 120fps.

This work establishes a new benchmark for low-latency, low-power event-based eye tracking on microcontrollers and paves the way for scalable deployment in wearable and always-on computing scenarios.  
\vspace{0.3cm}

% \section{Acknoledgement}
% We acknowledge the use of OpenAI’s ChatGPT in the preparation of this manuscript.

\bibliographystyle{unsrt}
\bibliography{bibliography}

\end{document}